\documentclass[twocolumn,superscriptaddress]{revtex4-1}
\usepackage{graphicx}%
\usepackage[T1]{fontenc}
\begin{document}
\title{Interlayer exciton dynamics in a dichalcogenide monolayer heterostructure}

\keywords{Transition-metal dichalcogenides, tungsten disulfide, excitons, trions, optical spectroscopy, valley dynamics}
\author{Philipp Nagler}
\affiliation{Institut f\"ur Experimentelle und Angewandte Physik,
	Universit\"at Regensburg, D-93040 Regensburg, Germany}
\author{Gerd Plechinger}
\affiliation{Institut f\"ur Experimentelle und Angewandte Physik,
	Universit\"at Regensburg, D-93040 Regensburg, Germany}

\author{Mariana V. Ballottin}
\affiliation{High Field Magnet Laboratory (HFML - EMFL), Radboud University, 6525 ED Nijmegen, The Netherlands}
\author{Anatolie Mitioglu}
\affiliation{High Field Magnet Laboratory (HFML - EMFL), Radboud University, 6525 ED Nijmegen, The Netherlands}
\author{Sebastian Meier}
\affiliation{Institut f\"ur Experimentelle und Angewandte Physik,
	Universit\"at Regensburg, D-93040 Regensburg, Germany}
\author{Nicola Paradiso}
\affiliation{Institut f\"ur Experimentelle und Angewandte Physik,
	Universit\"at Regensburg, D-93040 Regensburg, Germany}
\author{Christoph Strunk}
\affiliation{Institut f\"ur Experimentelle und Angewandte Physik,
	Universit\"at Regensburg, D-93040 Regensburg, Germany}
\author{Alexey Chernikov}
\affiliation{Institut f\"ur Experimentelle und Angewandte Physik,
	Universit\"at Regensburg, D-93040 Regensburg, Germany}
\author{Peter C. M. Christianen}
\affiliation{High Field Magnet Laboratory (HFML - EMFL), Radboud University, 6525 ED Nijmegen, The Netherlands}
\author{Christian Sch\"uller}
\affiliation{Institut f\"ur Experimentelle und Angewandte Physik,
	Universit\"at Regensburg, D-93040 Regensburg, Germany}
\author{Tobias Korn$^*$}
\affiliation{Institut f\"ur Experimentelle und Angewandte Physik,
	Universit\"at Regensburg, D-93040 Regensburg, Germany}
\email{tobias.korn@physik.uni-regensburg.de}
\begin{abstract}
In heterostructures consisting of different transition-metal dichalcogenide monolayers, a staggered band alignment can occur, leading to rapid charge separation of optically generated electron-hole pairs into opposite monolayers. These spatially separated electron-hole pairs are Coulomb-coupled and  form interlayer excitons. Here, we study these interlayer excitons in a heterostructure consisting of MoSe$_2$ and WSe$_2$ monolayers using photoluminescence spectroscopy. We observe a non-trivial temperature dependence of the linewidth and the peak energy of the interlayer exciton, including an unusually strong initial redshift of the transition with temperature, as well as a pronounced blueshift of the emission energy with increasing excitation power. By combining these observations with time-resolved photoluminescence measurements, we are able to explain the observed behavior as a combination of interlayer exciton diffusion and dipolar, repulsive  exciton-exciton interaction.
\end{abstract}
\maketitle
In recent years, two-dimensional crystal structures have garnered a lot of scientific attention. Using simple techniques such as mechanical exfoliation, a plethora of different materials is readily available as a two-dimensional sheet~\cite{Novoselov2005}, including large-gap insulators, superconductors, and semiconductors.  Due to quantum confinement effects, the electronic structure of these atomically thin layers can be very different from that of their corresponding bulk crystals. MoS$_2$ and related transition-metal dichalcogenides (TMDCs) such as WSe$_2$ and MoSe$_2$ are among the most promising systems: while they are indirect-gap semiconductors in the bulk, a transition to a direct band gap occurs as they are thinned down to a single layer~\cite{Lebegue2009, Splendiani2010, Mak2010}.
The peculiar band structure of the TMDC monolayers, combined with a strong spin-orbit interaction,  leads to a coupling of spin and valley degrees of freedom~\cite{Xiao2012,Xu2014a}, making these materials highly interesting for potential valleytronic applications.
Due to the strictly two-dimensional confinement of electrons and holes and the weak dielectric screening, excitons in these monolayer TMDCs are stable even at room temperature and exhibit  large binding energies of about 0.5~eV~\cite{Klots2014, Chernikov2014, Ugeda2014, Hanbicki2015, Poellmann2015}. While various TMDCs show qualitatively similar features, they are characterized by different absolute values of band gap and band offsets with respect to the vacuum level~\cite{Tongay13}. Therefore, several combinations of TMDCs were predicted to yield a staggered band alignment~\cite{Tongay13, Kosmider2013} when they are combined into a heterostructure, leading to spatial separation of optically generated electron-hole pairs. The development of various transfer techniques~\cite{Ponomarenko2011,Castellanos2014}  for building Van der Waals heterostructures~\cite{Geim2013} by stacking two-dimensional crystals made it possible to fabricate proof-of-concept devices such as light-emitting diodes and solar cells using TMDCs~\cite{Britnell,Furchi2014,Pospischil2016}, and to experimentally verify the predictions regarding band alignment for different TMDC combinations~\cite{Fang2014,Gong2014,Heo2015,Xu_NatComm15}. Remarkably,  photoluminescence (PL) spectra of TMDC heterostructures revealed the formation of interlayer excitons, where electrons and holes residing in different, adjacent TMDC layers are Coulomb-bound to each other, and can recombine radiatively.
Such interlayer excitons have already been studied in different material systems, e.g., in coupled quantum wells based on GaAs-AlGaAs heterostructures grown by molecular beam epitaxy (MBE). There, different aspects such as exciton-exciton correlations~\cite{Schindler2008,Rapaport_PRB09}, exciton condensation~\cite{PhysRevLett.74.1633,PhysRevLett.73.304,Snoke2002,High12}, macroscopically ordered exciton states~\cite{Butov02}  as well as manipulation of exciton diffusion using electric fields~\cite{Gartner2006} and exciton trapping~\cite{Hammack2006} were investigated. Interlayer excitons in TMDC heterostructures are characterized by binding energies which are expected to significantly exceed those in GaAs heterostructures~\cite{Wilsone1601832}, making them stable at room temperature and robust against dissociation in applied electric fields. Additionally, they  have long radiative lifetimes, in stark contrast to intralayer excitons in TMDC monolayers, where sub-picosecond radiative recombination can be observed~\cite{Poellmann2015,Marie16}. These long lifetimes potentially allow for cooling interlayer excitons to very low lattice temperatures, a prerequisite for exciton condensation. With this unique combination of properties, TMDC interlayer excitons are a highly promising novel platform for studying exciton-exciton interactions.

Here, we explore the properties of interlayer excitons and their interactions in a MoSe$_2$-WSe$_2$ heterostructure in a wide range of temperature and excitation power by means of photoluminescence spectroscopy. We observe pronounced and unusual energetic shifts of the interlayer exciton PL emission as a function of these experimental parameters. Combining the observations in time-integrated and time-resolved experiments, we are able to attribute the observed effects to a two-step process, considering exciton diffusion within an inhomogeneously broadened energy landscape and dipolar, repulsive exciton-exciton interaction.
\begin{figure}
\includegraphics*[width=\linewidth]{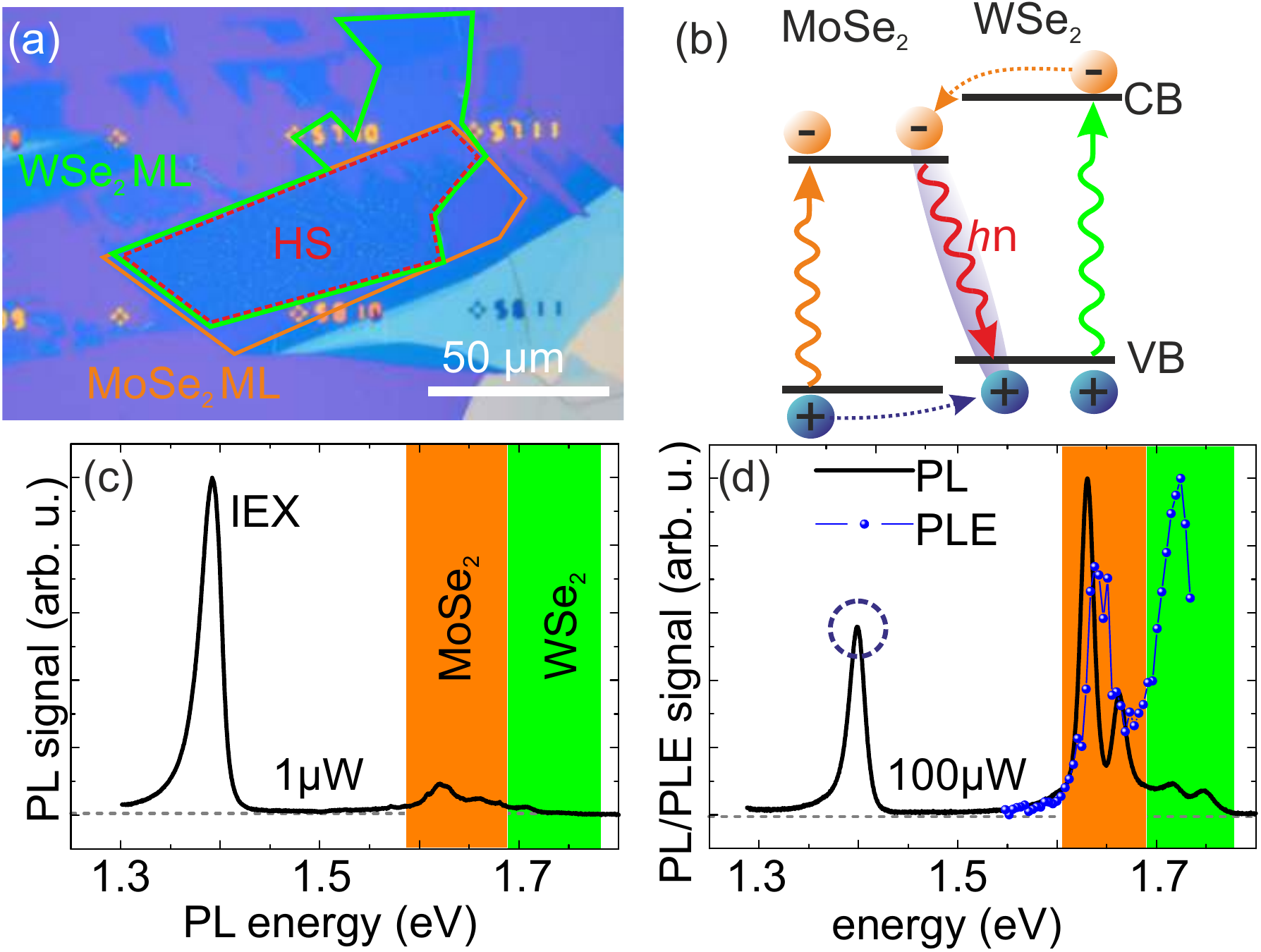}
\caption{\textbf{TMDC heterostructure characterization.} (a) Optical micrograph of sample structure. The colored outlines indicate the areas of the MoSe$_2$ (orange) and WSe$_2$ (green) MLs, as well as the heterostructure (HS, red dashed) formed in the overlap region of the two MLs.  (b) Schematic of the band alignment in the MoSe$_2$-WSe$_2$ HS. The staggered alignment of valence (VB) and conduction (CB) bands leads to spatial separation of electron-hole pairs after optical excitation of either of the two MLs. The spatially separated electron-hole pairs form interlayer excitons which decay radiatively. (c) PL spectrum of HS  measured at 4.5~K  with 1~$\mu$W excitation power. (d) PL spectrum of HS (black line) measured at 4.5~K with 100~$\mu$W excitation power and PLE spectrum (blue dots), where the PL intensity of the IEX emission (marked by the blue dotted circle) is recorded as a function of excitation laser energy. The orange and green bands in (c) and (d) indicate the spectral regions where PL emission from the MoSe$_2$ (orange) and WSe$_2$ (green) MLs is  observed.}
\label{fig1:Charact}
\end{figure}

Figure~\ref{fig1:Charact}(a) shows an optical microscope image of the studied sample structure, consisting of a WSe$_2$ flake transferred onto a MoSe$_2$ flake by deterministic transfer~\cite{Castellanos2014}. As indicated by the outlines of the  monolayer (ML) regions of the two flakes, two well-cleaved edges of the flakes are aligned parallel to each other during the fabrication process. In the heterostructure (HS) region formed in the overlap of the two MLs, small bubbles are clearly observable, indicating the coalescence of surface adsorbates trapped between the layers~\cite{Haigh} during annealing.

In low-temperature PL spectra  of the sample structure  we observe a pronounced emission at around 1.4~eV, in good agreement with previous reports on the interlayer exciton (IEX) emission energy in  MoSe$_2$-WSe$_2$ HS~\cite{Xu_NatComm15, Xu_Science16, Schaibley2016_2}. We note that all PL spectra shown in this manuscript are measured using pulsed excitation, see methods.  As Fig.~\ref{fig1:Charact}(c) shows, the PL emission of this low-energy peak is very pronounced compared to the intralayer exciton emission of the two constituent monolayers under weak excitation, due to very fast interlayer charge tunneling processes~\cite{Hong2014,Heinz_interlayer} which spatially separate electron-hole pairs generated by optical absorption in the individual monolayers. This charge separation is shown schematically in Fig.~\ref{fig1:Charact}(b). To support our assignment of the PL emission at 1.4~eV to the IEX, we perform photoluminescence excitation (PLE) measurements. Fig.~\ref{fig1:Charact}(d) shows the results of a PLE measurement, in which we record the PL emission intensity of the IEX as a function of the excitation laser wavelength. For comparison, the PLE data is superimposed onto a PL spectrum of the HS measured using a 100 times higher excitation power than in Fig.~\ref{fig1:Charact}(c), which yields a more pronounced intralayer exciton emission than weak excitation.  We clearly observe that IEX PL emission only emerges when the excitation laser energy is sufficiently high to match the PL emission of the MoSe$_2$ layer, and correspondingly, the minimum energy required for absorption in that layer. Additionally, the IEX PL emission is resonantly enhanced close to the PL maxima of both, the MoSe$_2$ and the WSe$_2$ layers, due to the corresponding absorption maxima in the layers. The PLE spectrum thus strongly supports the interlayer nature of the IEX emission peak: for the majority of alternative scenarios, including very rapid capture of an intra-layer exciton to a defect in either of the two individual monolayers, we would rather expect to observe resonances related only to the corresponding layer, not to both of the MLs.
\begin{figure}
\includegraphics*[width=\linewidth]{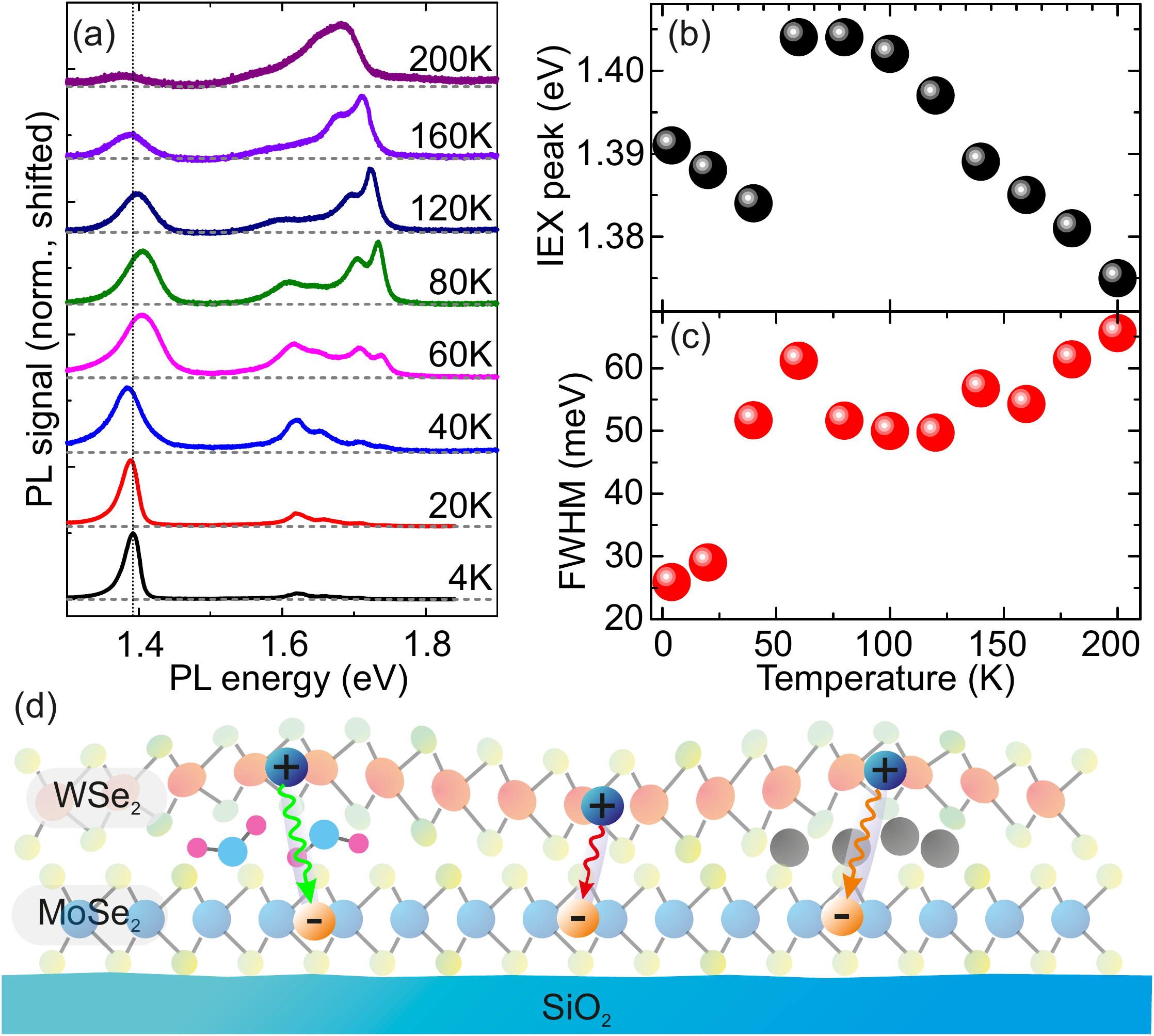}
\caption{\textbf{Photoluminescence of interlayer excitons: Temperature dependence.} (a) Normalized PL spectra of HS measured as a function of temperature using fixed excitation power. The vertical line  marks the peak position of the IEX at low temperature and serves as guide to the eye. (b) IEX peak position and (c) IEX spectral linewidth (FWHM) as a function of temperature. (d) Schematic of HS with adsorbates between the two TMDC layers modulating the interlayer distance and, correspondingly, the exciton binding and IEX PL emission energy.}
\label{fig2:T-PL}
\end{figure}

To explore the stability of interlayer excitons and exciton-exciton interaction effects, we now turn to the dependence of the time-integrated IEX PL emission on temperature and excitation density. Figure~\ref{fig2:T-PL}(a) shows a series of PL spectra of our HS sample at different temperatures. Several striking features are immediately apparent: the IEX PL emission is observable in the whole temperature range investigated in this series, demonstrating the stability of the IEX against thermal dissociation. At low temperatures, the intralayer exciton  PL emission of MoSe$_2$ and WSe$_2$ is strongly suppressed, and increases, relative to the IEX emission intensity, as a function of temperature, between 4~K and 60~K. MoSe$_2$ and WSe$_2$ show markedly different behavior as the temperature increases further, with the PL yield of WSe$_2$ continuously increasing up to 200~K, while the MoSe$_2$ yield decreases with increasing temperature, albeit not as rapidly as that of the IEX (see supplementary note 3). The unusual increase of the WSe$_2$ PL yield, which was previously reported by several groups~\cite{Arora15,Urbaszek_MoWSe,Heinz_DarkEx_PRL15} is a direct consequence of the conduction-band spin splitting in Tungsten-based TMDCs. More remarkably, we note a very unusual behavior of the spectral position  of the IEX peak as a function of temperature (see Fig.~\ref{fig2:T-PL}(b)). Between 4.5~K and 40~K,  the IEX peak  redshifts by about 10~meV with increasing temperature, then increases its energy by more than 20~meV as the temperature is raised from 40~K to 60~K. In the same temperature range, both intralayer exciton and trion emission peaks show only a weak, continuous redshift of about 5~meV, which is well-described by the Varshni formula (see supplementary note 1), and continues throughout the whole temperature range investigated here. Thus, neither the large IEX redshift between 4.5~K and 40~K, which significantly exceeds the redshift of the intralayer excitons, nor the large blueshift between 40~K and 60~K, can be explained by changes related to the transition energies of the constituent MLs of the HS.  The IEX peak position has a plateau between 60~K and 80~K, and red-shifts as the temperature is increased further. This redshift at higher temperatures has a similar slope as that of the intralayer exciton emission peaks of MoSe$_2$ and WSe$_2$  (see supplementary note 1). We also observe an unusual temperature dependence of the spectral linewidth of the IEX peak (see Fig.~\ref{fig2:T-PL}(c)). It almost doubles as the temperature is increased from 4~K to 40~K, reaches a maximum at 60~K and then remains nearly constant between 80~K and 120~K. As the temperature is increased further, the IEX linewidth increases monotonously.  By contrast, the intralayer exciton emission  of MoSe$_2$ and WSe$_2$ shows a weaker, continuous broadening with temperature in the whole temperature range investigated here (see supplementary note 1),  induced by increased exciton-phonon scattering with increasing temperatures~\cite{Selig2016}.

A qualitatively similar, nonmonotonous behavior of excitonic emission energy as a function of temperature, typically referred to as an 's-shape', coupled with a rapid increase of the emission linewidth, has previously been observed in disordered bulk semiconductors and quantum wells~\cite{Skolnick86, doi:10.1063/1.2058192,Rubel2007285,doi:10.1063/1.3374884,doi:10.1063/1.3580773}. In these systems, excitons can be trapped in local minima of the disorder potential at low temperatures. With increasing temperature, exciton diffusivity via phonon-assisted hopping increases, so that excitons can diffuse to energetically lower trap states, initially leading to a pronounced redshift within a small temperature range, combined with a broadening of the PL linewidth as a larger range of localized trap states becomes accessible~\cite{PhysRevB.58.13081}. A possible source of disorder for interlayer excitons is related to the layer-by-layer fabrication of our heterostructures, as illustrated in Fig.~\ref{fig2:T-PL}(d). Our fabrication procedure  leads to the inclusion of adsorbates in between the layers, so that the interlayer distance  varies a function of position, with a minimum interlayer distance for regions where there are no adsorbates between the two monolayers. This variation of the interlayer distance also modifies the electron-hole spacing and, correspondingly, the IEX binding energy, with a maximum binding energy, and corresponding minimum total energy $E_{Min}$ limited by the smallest interlayer distance. Additional disorder and localized trap states are potentially introduced by charged adsorbates above or below the heterostructure. Temperature-activated interlayer exciton diffusion within this inhomogeneous potential landscape will therefore lead to an initial, pronounced redshift of the PL energy, as local and global potential minima become accessible. However, as we will discuss below, in our structures we also need to consider a pronounced temperature dependence of the IEX lifetime, and density-dependent exciton-exciton-interaction effects which modify the PL emission energy.

\begin{figure}
\includegraphics*[width=\linewidth]{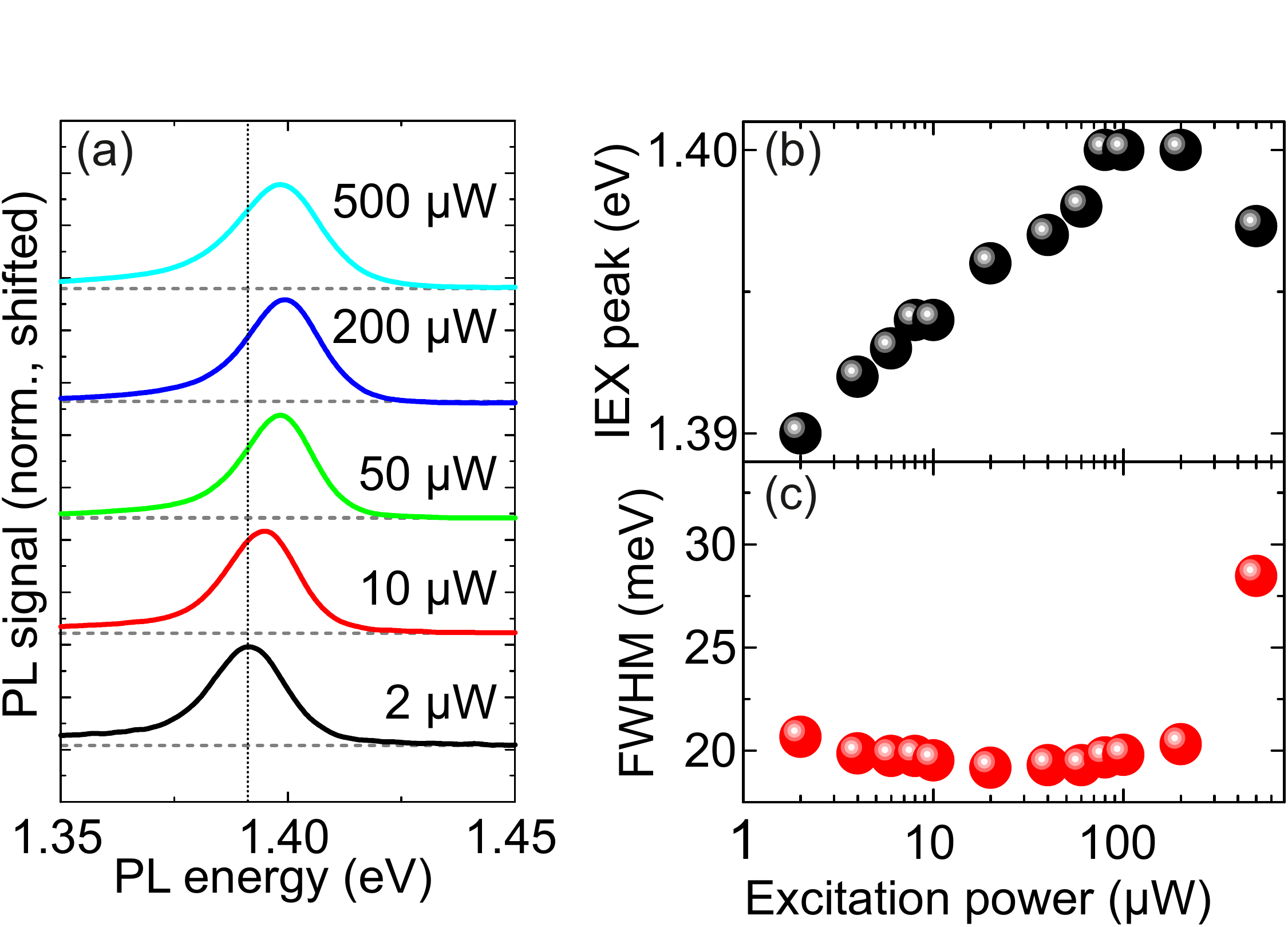}
\caption{\textbf{Photoluminescence of interlayer excitons: Power dependence.} (a) PL spectra of HS measured at 4.5~K as a function of excitation power. The vertical line marks the peak position of the IEX at low power and serves as guide to the eye. (b) IEX peak position and (c) IEX FWHM as a function of excitation power.}
\label{fig3:Power-PL}
\end{figure}

Power-dependent PL measurements at 4.5~K (Fig.~\ref{fig3:Power-PL}(a)) also indicate an unusual behavior of the IEX peak emission. For a wide range of excitation powers, we find that the peak position blueshifts with increasing power by up to 10~meV, as Fig.~\ref{fig3:Power-PL}(b) shows. The slope of this blueshift is sub-linear. Remarkably, the IEX linewidth (Fig.~\ref{fig3:Power-PL}(c)) remains almost constant in the large power range (spanning almost 2 orders of magnitude) in which we observe this continuous blueshift. This blueshift with increasing excitation power is reproduced in a second measurement series using continuous-wave excitation instead of pulsed excitation (see supplementary note 2), and qualitatively matches a previous study~\cite{Xu_NatComm15}.
As we increase the laser power further, the IEX blueshift saturates, then the IEX peak energy decreases again. In this high-power range, we also see an increase of the IEX spectral linewidth. By contrast, the intralayer exciton emission of MoSe$_2$ shifts by less than 1~meV in the whole power range investigated here. Clearly, the observed blueshift cannot be explained by laser-induced heating, as we observe the opposite effects (strong redshift and linewidth increase) with increasing sample temperature.

To determine the mechanism responsible for the  unusual power dependence of the IEX PL emission, we turn to time-resolved photoluminescence (TRPL) measurements.
\begin{figure}
\includegraphics*[width=\linewidth]{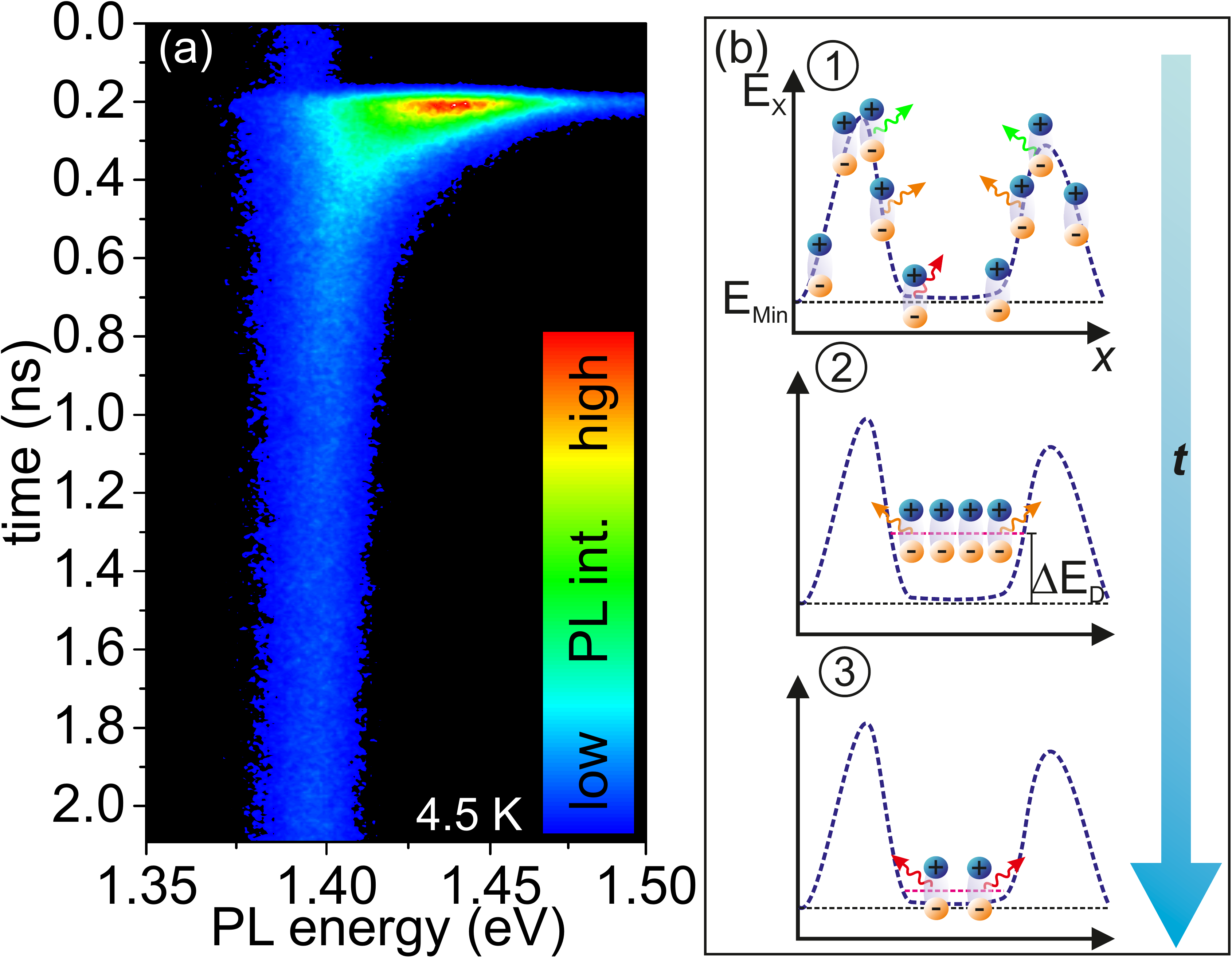}
\caption{\textbf{Dynamics of interlayer exciton photoluminescence} (a) False-color plot of TRPL spectrum  measured for 4.5~K.  (b) Schematic of interlayer exciton dynamics as a function of time: (1) interlayer excitons are generated within the  inhomogeneous potential landscape of the HS, where the IEX total energy $E_{X}$ varies as a function of position $x$ and the minimum energy $E_{Min}$ is given by the minimum interlayer distance. They diffuse to the energetically favorable local minima of the potential. Spectrally broad IEX PL emission occurs during this diffusion process, as excitons at different positions within the potential landscape recombine. (2) Interlayer excitons are trapped within the potential minima. Repulsive dipole-dipole interaction blueshifts the exciton energy by a density-dependent term $\Delta E_{D}$, leading to spectrally narrow, blueshifted IEX PL emission. (3) As the IEX density within the potential minima is reduced, the IEX PL emission redshifts towards the limit $E_{Min}$.}
\label{fig4:TRPL-mechanism}
\end{figure}
Figure~\ref{fig4:TRPL-mechanism}(a) shows a false-color plot of the IEX PL intensity as a function of time and energy, measured at 4.5~K using a streak camera system. We observe that the PL directly after excitation is blue-shifted and spectrally broadened compared to the emission at later times. We clearly see a very long-lived, spectrally narrow 'tail' of the PL emission, exceeding the measurement window of 2~ns, and even some IEX PL emission 'before' the arrival of the laser pump pulse, indicating that the IEX PL lifetime exceeds the 12.5~ns time window between subsequent excitation pulses. PL lifetime measurements using a different setup show a biexponential decay behavior of the PL emission at 4.5~K (see supplementary note 4), yielding a value of 138~ns for the slow component.
We can qualitatively explain these photoluminescence dynamics, as depicted in Fig.~\ref{fig4:TRPL-mechanism}(b), by considering the diffusion of the IEX in an inhomogeneous potential landscape.
After pulsed excitation, interlayer excitons form throughout the whole HS, and diffuse to the local potential minima. During this diffusion process, a part of the interlayer excitons recombines radiatively, leading to a spectrally broad PL emission that is, on average, at a higher energy than $E_{Min}$. The remaining interlayer excitons are trapped within the local potential minima, cool down towards the lattice temperature via phonon emission, and the inhomogeneous broadening of the PL emission is reduced. However, we also need to consider the fact that the interlayer excitons carry a permanent electrical dipole moment due to the separation of charges into the different TMDC layers, which leads to a repulsive, dipolar exciton-exciton interaction. This repulsive interaction manifests itself as a density-dependent blueshift $\Delta E_{D}$ of the total exciton energy~\cite{Butov99,Rapaport_PRB09}. As the exciton density decreases, $\Delta E_{D}$ is reduced, and the IEX emission energy is decreased towards $E_{Min}$.

\begin{figure}
\includegraphics*[width=\linewidth]{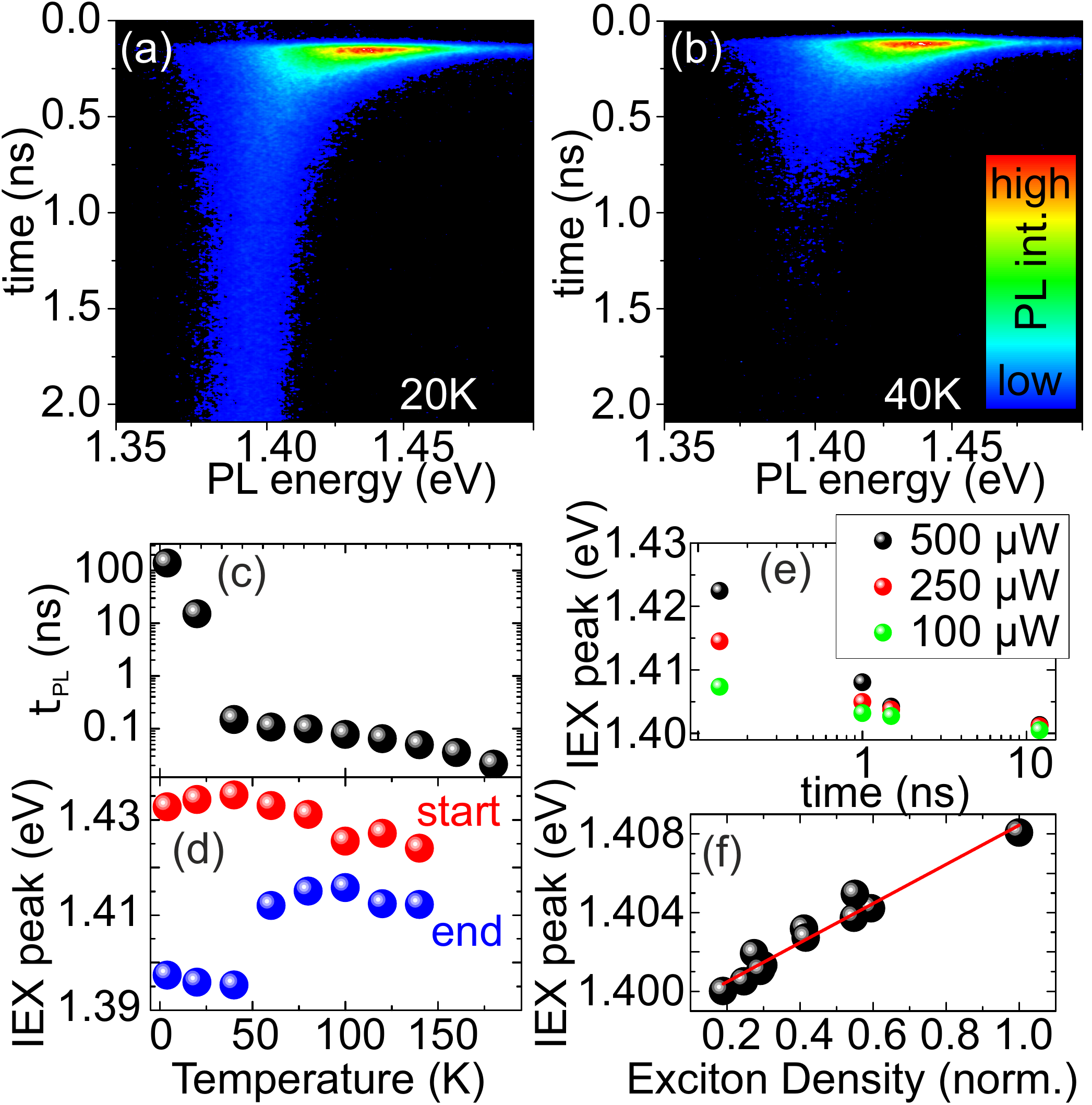}
\caption{\textbf{Power- and temperature-dependent time-resolved photoluminescence.} (a) and (b)  False-color plots of TRPL spectra  measured for 20~K (a) and 40~K (b). (c) PL lifetime of IEX as a function of temperature. (d) IEX peak position at the start and end of a TRPL trace as a function of temperature. (e) IEX peak position as a function of time, measured at 4.5~K for 3 different excitation power values. (f) IEX peak position as a function of normalized exciton density, extracted from power-dependent TRPL traces.}
\label{fig5:TRPL}
\end{figure}

To further elucidate the energetic shifts of the IEX observed in the time-integrated PL measurement series, we now focus on the temperature and power dependence of the IEX PL dynamics. Figure~\ref{fig5:TRPL}(a) and (b) show false-color plots of the IEX PL intensity as a function of time and energy, measured at 20~K and 40~K, respectively. Both spectra show a qualitatively similar behavior as observed at 4.5~K: the PL after excitation is blue-shifted and spectrally broadened. However, the total lifetime of the PL emission is significantly shorter at 40~K. For these traces and those measured at higher temperatures, we are able to directly extract the PL lifetime $\tau_{PL}$ using the streak camera system. This is plotted in Fig.~\ref{fig5:TRPL}(c), showing a drastic drop of the PL lifetime with increasing temperature. We find that this drastic reduction of the PL lifetime by more than two orders of magnitude between 4.5~K and 40~K is not accompanied by a correspondingly drastic reduction of PL yield (see supplementary note 3 for the temperature dependence of IEX and intralayer PL yield), indicating that it is not purely related to a thermally activated nonradiative decay channel or Auger-type recombination. The microscopic origin of this drastic reduction of $\tau_{PL}$ deserves further study. One possible explanation is an imperfect alignment of the conduction- and valence-band extrema in the adjacent TMDC layers in \emph{k} space. In this case, momentum conservation requires a phonon-assisted process for  radiative recombination of the IEX.
To further analyze the TRPL dynamics, we determine the IEX PL peak position at the start and close to the end of the PL lifetime. For this we define temporal windows with a width of 100~ps, centered at the onset of the IEX PL emission and at a time after excitation where the PL has decayed almost fully, but to a level that still allows us to extract the peak position. Due to the reduction of $\tau_{PL}$ with increasing temperature, this end time shifts to smaller values with increasing temperature. The results of this analysis are given in Fig.~\ref{fig5:TRPL}(d). We clearly see that for temperatures between 4.5~K and 40~K, there is a large energy difference (more than 30~meV) between the IEX emission at the onset of PL and the redshifted emission at late times. This difference is drastically reduced to less than 20~meV as the temperature is raised to 60~K, and further decreases with increasing temperature, indicating that the reduction of PL lifetime with temperature limits the available time window for energy relaxation of interlayer excitons via diffusion. These insights into the IEX PL dynamics allow us to understand the temperature-dependent behavior of the time-integrated PL emission shown in Fig.~\ref{fig2:T-PL}(a-c): as the sample temperature is increased from 4.5~K to 40~K, the diffusivity of interlayer excitons increases, as discussed above, so that lower-energy states within the inhomogeneous potential landscape become accessible during exciton diffusion. Additionally, however, the temperature increase drastically decreases the IEX lifetime, leading to a reduction of the time-averaged interlayer exciton density for  constant excitation power. Therefore, the effect of repulsive exciton-exciton interaction is reduced, also corresponding to a redshift of the time-integrated PL emission. As the temperature is increased from 40~K to 60~K and more, the PL lifetime becomes so short that on average, interlayer excitons are no longer able to diffuse into the local energy minima. Thus, the PL spectral linewidth increases, as observed in Fig.~\ref{fig2:T-PL}(c), due to PL emission from energetically unfavorable regions of the potential landscape, and the spectral weight of the emission shifts to higher energy. As the temperature is increased even further, bandgap reduction of the constituent monolayers leads to the continuous redshift of the IEX emission.

In order to study the influence of interlayer exciton density in more detail, we focus on the excitation power dependence of the TRPL traces. In this measurement series, the sample temperature was kept fixed at 4.5~K. In Fig.~\ref{fig5:TRPL}(e), we analyze the IEX peak position as a function of time for 3 different excitation powers. We clearly see a qualitatively similar behavior for all excitation powers: the PL emission redshifts as a function of time. Remarkably, we find that the IEX peak position strongly depends on the excitation power throughout the whole time window investigated here.
To analyze the dependence of the IEX peak position on exciton density, we make the assumption that at sufficiently long time delay after excitation to allow for exciton energy relaxation (here, we consider only delay times above 800~ps after excitation), the PL emission intensity will be proportional to the exciton density. Using this assumption, we can relate PL spectra extracted from TRPL traces measured under several different excitation powers at different time delays to each other, by using the integrated PL intensity as a measure of exciton density (see supplementary note 6). The results of this analysis are depicted in Fig.~\ref{fig5:TRPL}(f). We  see that the IEX peak position has a blueshift that increases linearly with the exciton density. This linear dependence, which is expected from a basic mean-field approximation~\cite{Rapaport_PRB09}, strongly supports our assessment that the observed blueshift is due to dipolar exciton-exciton interaction.

In the time-integrated, power-dependent PL experiments (see Fig.~\ref{fig3:Power-PL}(a-c)), the observed blueshift with increasing excitation density is caused by the repulsive exciton-exciton interaction. The sublinear slope of the blueshift indicates that the effective IEX PL lifetime \emph{decreases} with increasing power. This is due to the increase of the electron-hole pair density in both TMDC layers which allows for intralayer exciton recombination and reduces the efficiency of IEX formation. This interpretation is confirmed by  power-dependent time-resolved PL traces (see supplementary note 4) and by tracking the time-integrated PL intensity of IEX and intralayer exciton emission, where intralayer exciton emission increases its relative intensity with respect to IEX emission with increasing excitation power (see supplementary note 5).
This reduction of the IEX PL lifetime also limits the IEX density that can be trapped within the potential minima using pulsed excitation, leading to a saturation of the PL blueshift. This is accompanied by an increased IEX PL linewidth due to emission from energetically unfavorable regions of the potential landscape. Remarkably, by comparing the power-dependent measurement series using weak, continuous-wave (cw) excitation with the time-resolved PL measurements under intense pulsed excitation, we find that the tuning range for IEX PL emission due to the dipolar-repulsion-induced blueshift spans at least 20~meV. A simple estimation using the plate capacitor formula~\cite{Butov99} for calculating the maximum interlayer exciton density $n_{IEX}$ based on this blueshift yields a lower boundary  value of $n_{IEX}=4\cdot10^{10}$~cm$^{-2}$ (see supplementary note 7).

In conclusion, we have explored interlayer exciton photoluminescence in a TMDC heterostructure as a function of temperature and excitation power. We observe pronounced and unusual energetic shifts of the interlayer exciton PL emission as a function of these experimental parameters. By combining these results with time-resolved photoluminescence measurements which track the interlayer exciton recombination dynamics, we are able to  explain the observed behavior based on a combination of two processes: interlayer exciton diffusion within the inhomogeneous potential landscape, and dipolar repulsive exciton-exciton interaction. Our results are of particular importance for the understanding of the interlayer-exciton physics with respect to localization phenomena and further establish TMDC heterostructures as a robust platform for studying exciton-exciton interactions. With their unique combination of large binding energies, long lifetimes and the potential to explore the valley degree of freedom, TMDC interlayer excitons are likely to re-invigorate research in this field.
\subsection*{Acknowledgements}
The authors gratefully acknowledge fruitful discussions with J. Kunstmann. Financial support by the DFG via GRK 1570, KO3612/1-1,CH 1672/1-1 and SFB 689, as well as support of HFML-RU/FOM, member of the European Magnetic Field Laboratory (EMFL), is also gratefully acknowledged.
\section*{Methods}
\subsection{Sample preparation}
The heterostructure sample was fabricated by means of a deterministic transfer process~\cite{Castellanos2014}. For this, we initially exfoliated MoSe$_2$ and  WSe$_2$ flakes from bulk crystals (HQ graphene) onto intermediate polydimethylsiloxane(PDMS) substrates. Monolayer regions of these flakes were identified via optical microscopy. Then, we first transferred a MoSe$_2$ flake onto the target substrate, a silicon wafer piece covered with an SiO$_2$ layer and pre-defined metal markers. Subsequently, the WSe$_2$ flake was transferred on top of the MoSe$_2$. During this transfer, we carefully aligned well-cleaved edges of the monolayer parts of the two flakes to yield crystallographic alignment of the layers. Subsequent to the transfer, the sample was annealed in vacuum at a temperature of 150$^o$C for 5 hours to improve interlayer coupling~\cite{Tongay2014}.
\subsection{Optical spectroscopy}
Time-integrated photoluminescence (PL) measurements were performed in a self-built confocal microscope setup. A frequency-doubled pulsed fiber laser system (pulse length (FWHM) 180~fs,  pulse repetition rate 80~MHz) tuned to a central wavelength of 560~nm was used as excitation source,  coupled into a 100x microscope objective and focussed to a spot diameter of less than 1~micron on the sample surface. The PL from the sample was collected using the same objective and coupled into a grating spectrometer, where it was detected using a charge-coupled device (CCD) sensor. The sample was mounted on the cold finger of a small He-flow cryostat and scanned beneath the microscope objective. Time-resolved PL (TRPL) measurements were performed using the same microscope setup.  The PL emission was detected using a streak camera system coupled to the grating spectrometer and electronically  synchronized with the pulsed laser system. The temporal resolution of this setup is below 10~ps.
Photoluminescence excitation (PLE) measurements were performed in a similar self-built confocal microscope setup, with the sample mounted in a He-flow cryostat. A tunable cw Titanium:sapphire laser system  was used as  excitation source, the PL emitted by the sample was coupled into a grating spectrometer and detected using a liquid-nitrogen-cooled CCD. For the PLE measurements, the laser wavelength was varied in 2~nm increments, while the laser power was kept constant at 100~$\mu$W.
\onecolumngrid
\appendix
\subsection*{Supplementary note 1: temperature dependence of intralayer PL emission energy and linewidth}
\begin{figure}[b]
\includegraphics*[width=0.4\linewidth]{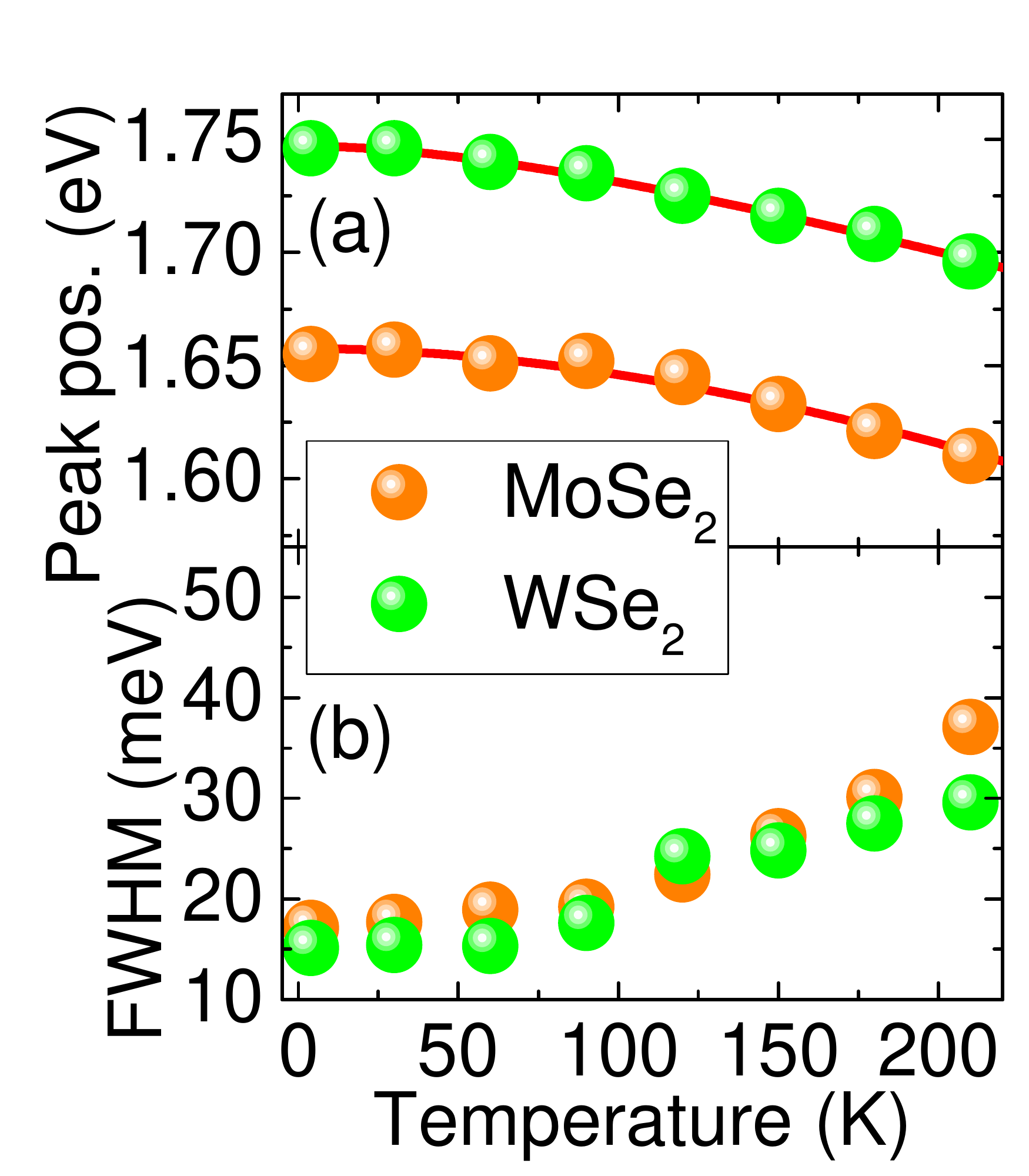}
\caption{\textbf{Temperature dependence of intralayer exciton PL peak position and linewidth.} (a) Exciton PL peak positions for MoSe$_2$ (orange dots) and WSe$_2$ (green dots) monolayers as a function of temperature. The solid red lines indicates fits to the data using the Varshni formula. (b) Spectral linewidth (FWHM) of exciton PL emission peaks for MoSe$_2$ (orange dots) and WSe$_2$ (green dots) monolayers as a function of temperature.}
\label{figSupp:T-dep}
\end{figure}

Figure~\ref{figSupp:T-dep}(a) shows the PL peak positions for the  exciton emission of MoSe$_2$ and WSe$_2$ monolayers as a function of temperature. To simplify data acquisition, the PL spectra from which these data were extracted were acquired on isolated MoSe$_2$ and WSe$_2$ monolayers, not on the heterostructure region, where the intralayer PL emission is significantly suppressed at low temperatures. However, in the heterostructure region, the intralayer PL emission shows the same behavior. The observed redshift with increasing temperature is well-described by the Varshni formula~\cite{Varshni1967}, as the fits to the data indicate. The absolute value of the redshift (about 30~meV between 100~K and 200~K for both materials) is in good quantitative agreement with the redshift observed for the interlayer exciton peak in the same temperature range, indicating a common origin, temperature-induced bandgap reduction. The spectral linewidth of the intralayer PL emission as a function temperature is depicted in Figure~\ref{figSupp:T-dep}(b). Here, we clearly observe a monotonic increase of the linewidth with increasing temperature due to exciton-phonon interaction.
\subsection*{Supplementary note 2: power dependence of interlayer exciton PL position under cw excitation}
\begin{figure}
\includegraphics*[width=0.4\linewidth]{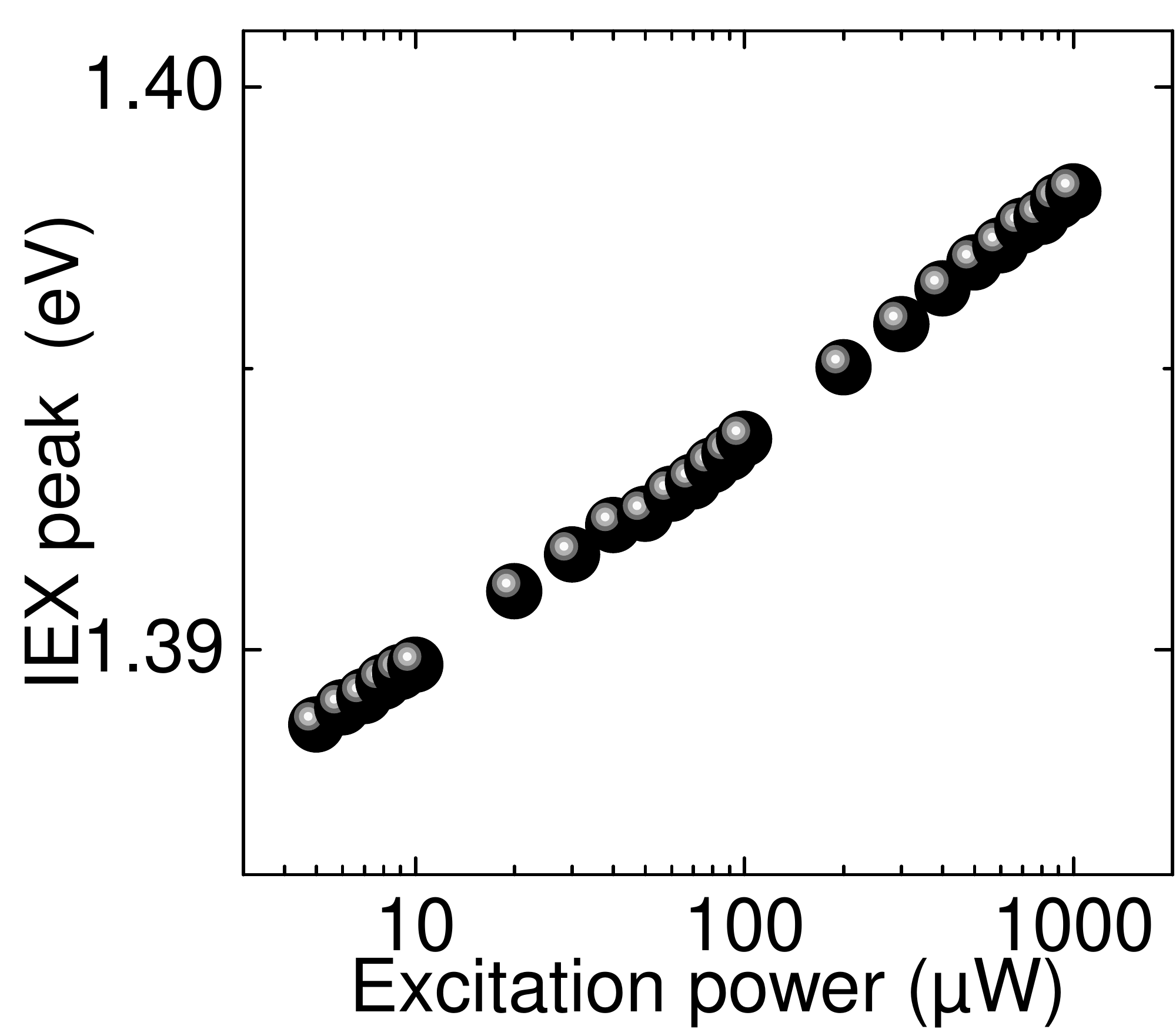}
\caption{\textbf{Interlayer exciton peak position as a function of power under cw excitation.} IEX PL peak position (black dots) as a function of laser power extracted from PL spectra measured at 4.5~K using cw laser excitation at a wavelength of 715~nm.}
\label{figSupp:cw_PeakPos_Power}
\end{figure}

Figure~\ref{figSupp:cw_PeakPos_Power} shows the IEX PL peak position as a function of excitation power under continuous-wave excitation. This measurement series was performed in the setup that was also used for the PLE measurements shown in the main text. In this series, the sample temperature was kept fixed at 4.5~K, and the excitation laser was tuned to a wavelength of 715~nm. The IEX PL peak position was extracted from the spectra by fitting a Gaussian. We clearly see a blueshift of the peak position as a function of excitation power, spanning a similar range of about 10~meV as observed in the power-dependent measurements under pulsed excitation shown in the main text, and also showing a sublinear slope of the blueshift as a function of power.  Remarkably, in contrast to the measurements shown in the main text, we do not observe a saturation of the blueshift in this measurement series.
\subsection*{Supplementary note 3: temperature dependence of interlayer and intralayer exciton PL yield}
\begin{figure}
\includegraphics*[width=0.4\linewidth]{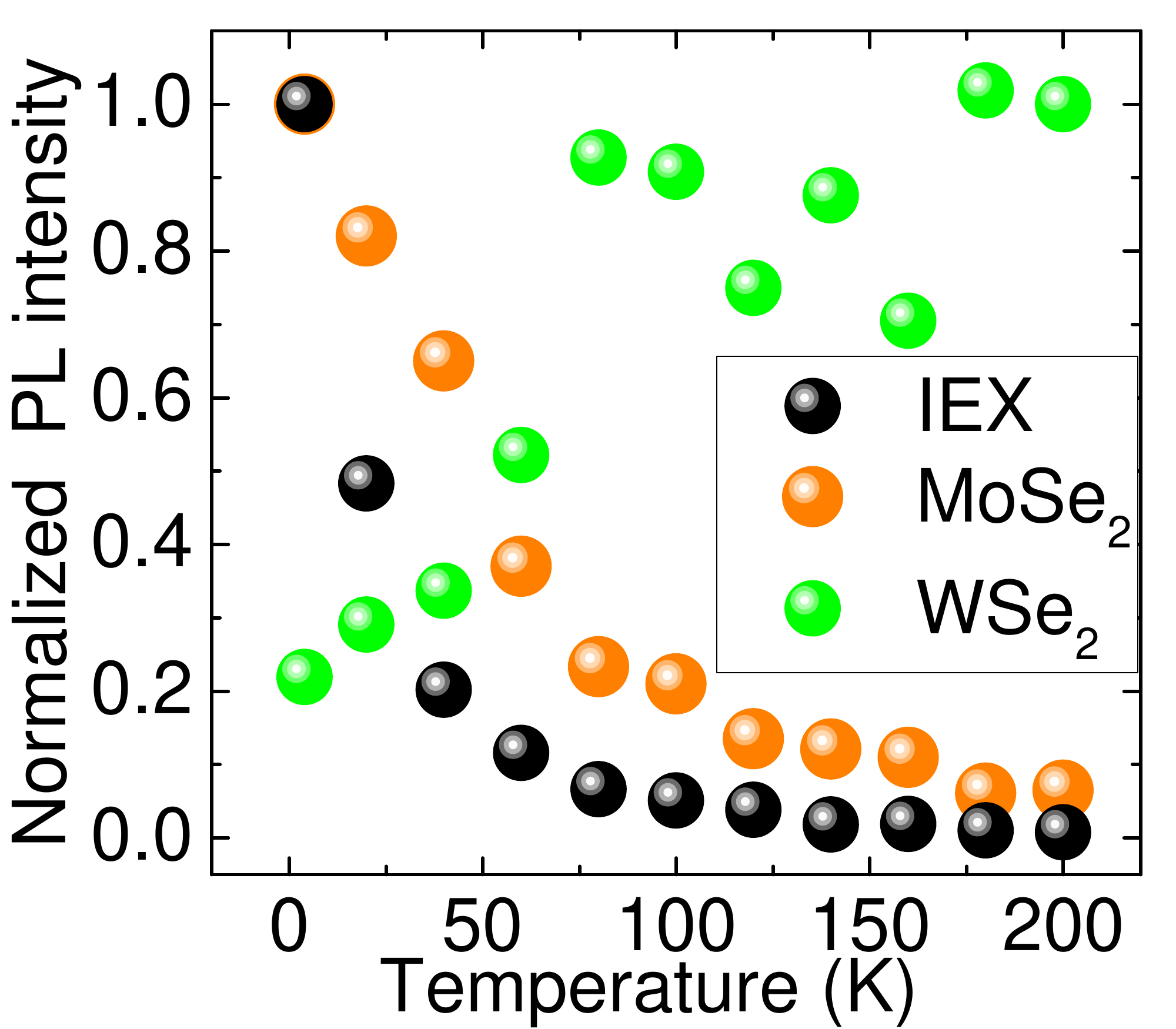}
\caption{\textbf{PL intensity of inter- and intralayer exciton emission as a function of temperature.} PL intensity of IEX  (black dots),  intralayer MoSe$_2$ (orange dots) and WSe$_2$ (green dots) as a function of temperature under cw excitation.}
\label{figSupp:intens_T-dep}
\end{figure}

Figure~\ref{figSupp:intens_T-dep} shows the normalized PL intensity of interlayer exciton and the intralayer MoSe$_2$ and WSe$_2$ emission as a function of temperature. For  MoSe$_2$ and WSe$_2$, trion and neutral exciton peaks were each fitted with a Gaussian, and the total PL intensity was calculated as the sum of the two areas. The PL intensities for IEX, MoSe$_2$ and WSe$_2$ were then normalized to the maximum PL intensity within the investigated temperature range. We see that the PL intensities of IEX and MoSe$_2$ monotonously decrease with increasing temperature, while the WSe$_2$ emission increases with increasing temperature due to thermal activation of the A exciton transition.
\subsection*{Supplementary note 4: interlayer exciton PL lifetime at low temperatures}
\begin{figure}
\includegraphics*[width=\linewidth]{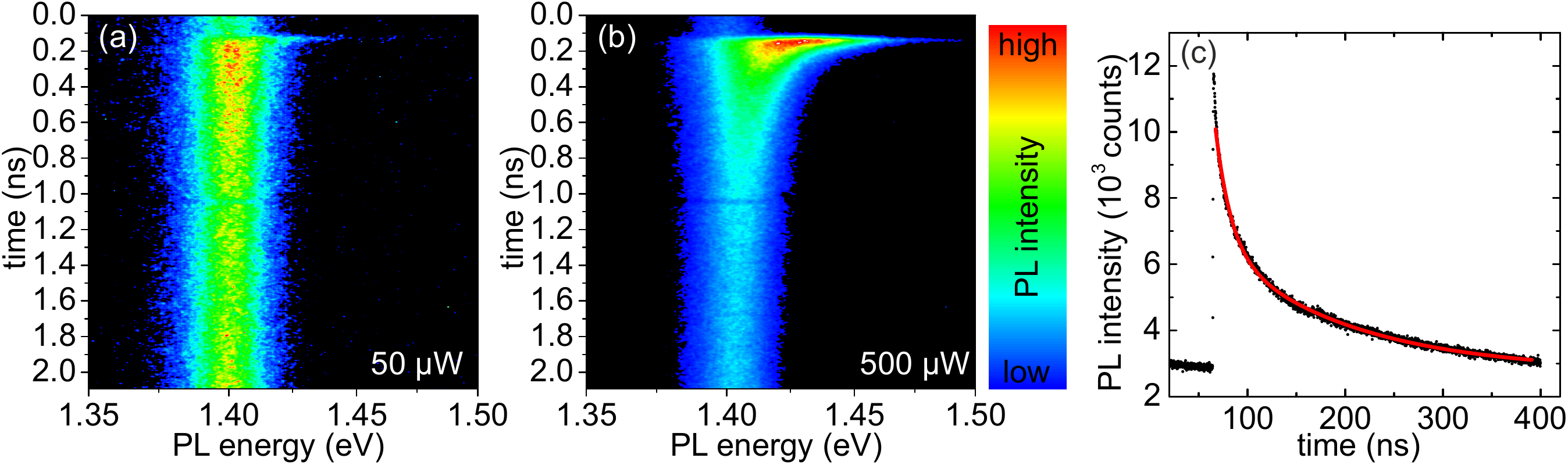}
\caption{\textbf{Interlayer exciton PL dynamics at low temperatures.} (a) and (b)  False-color plots of TRPL spectra  measured at 4.5~K for 50~$\mu$W (a) and 500~$\mu$W (b) excitation power. (c) PL intensity of IEX  (black dots) as a function of time, measured at 4.5~K using pulsed diode laser and avalanche photodiode. The red solid line is a biexponential decay function fitted to the data.}
\label{figSupp:PL_trace}
\end{figure}

At temperatures below 20~K, the PL lifetime of the interlayer exciton significantly exceeds the time window accessible in our streak camera measurements, which is given by the the inverse of the laser repetition rate of 80~MHz. This is clearly visible in the false color plots shown in Figure~\ref{figSupp:PL_trace} (a) and (b), where pronounced IEX PL is observed for time delays 'before' the arrival of an excitation pulse, indicating a background interlayer exciton density which stems from the superposition of previous laser pulses. Remarkably, direct comparison of the false color plots vividly shows that for large excitation power (Fig.~\ref{figSupp:PL_trace} (b)), the IEX PL originating from a single pulse dominates the false color plot, showing a pronounced redshift during the first 500~ps after arrival of the excitation pulse, combined with a significant decay of the PL intensity. By contrast, for weak pumping (Fig.~\ref{figSupp:PL_trace} (a)), the PL emission which stems from the superposition of previous laser pulses is of comparable intensity as that created by a single pulse. This indicates that the effective PL lifetime of the IEX emission at low temperatures is power-dependent, and difficult to accurately determine due to the short time window available in the streak camera system. Thus, we utilized an alternative micro-PL setup, in which we used a pulsed diode laser system (pulse length 80~ps, wavelength 690~nm, average power 18~$\mu$W, repetition rate 2.5~MHz) synchronized to an avalanche photodiode (APD). The PL emitted from the sample under this excitation was coupled into a monochromator set to the IEX emission energy, and the APD was used to detect the PL intensity as a function of time. Figure~\ref{figSupp:PL_trace} shows a PL trace detected in this setup. We utilized a biexponential fit function to extract the PL lifetime from this trace, which yields values of 16~$\pm 0.2$~ns and 138~$\pm 2$~ns for the fast and slow components of the decay, respectively.
\subsection*{Supplementary note 5: power dependence of interlayer and intralayer exciton PL yield}
\begin{figure}
\includegraphics*[width=0.4\linewidth]{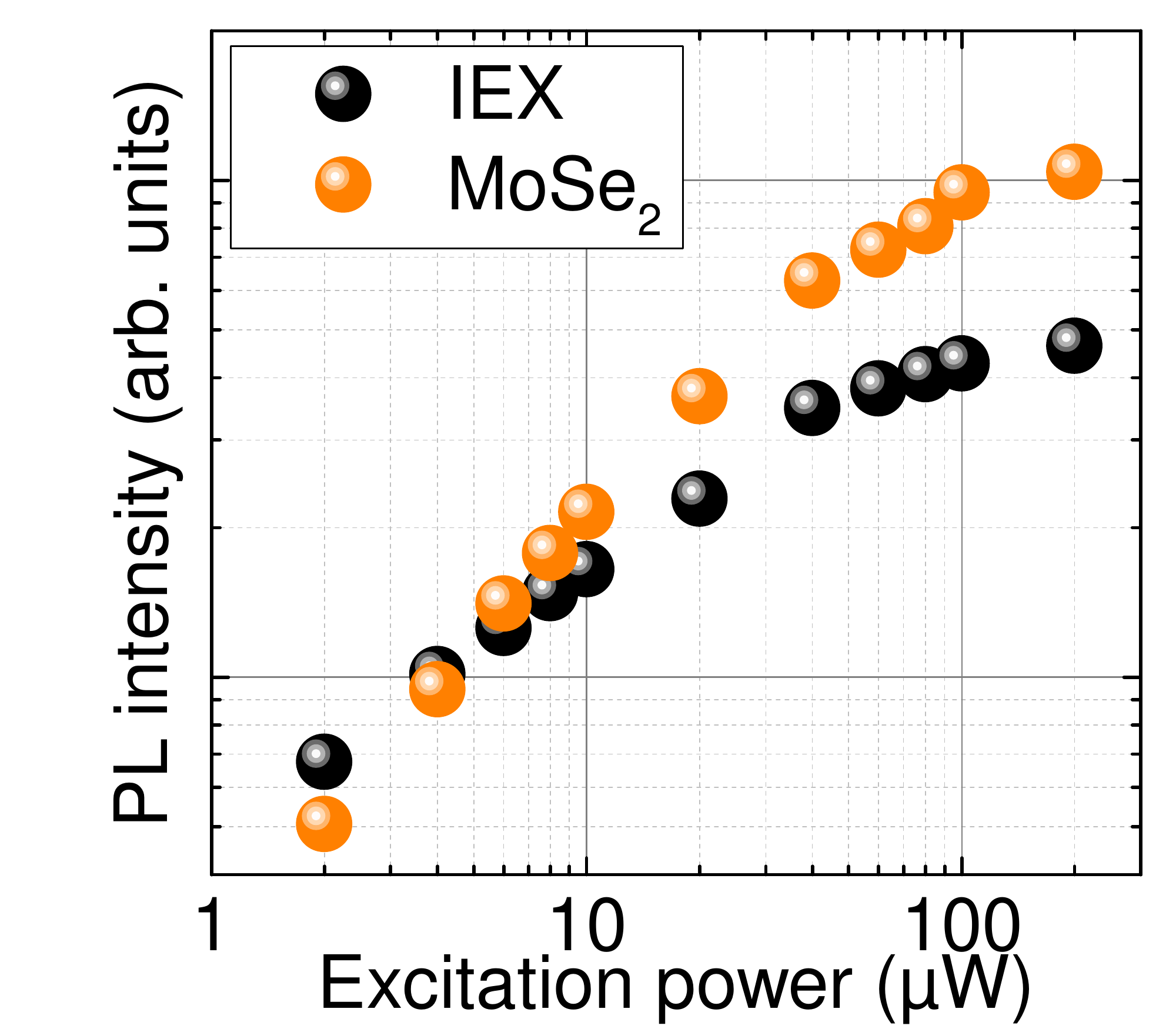}
\caption{\textbf{PL intensity of inter- and intralayer exciton emission as a function of excitation power.} PL intensity of IEX  (black dots) and intralayer MoSe$_2$ emission (orange dots) as a function of excitation power, measured at 4.5~K.}
\label{figSupp:P-dep}
\end{figure}

Figure~\ref{figSupp:P-dep} shows the PL intensity of  interlayer exciton and intralayer MoSe$_2$ emission as a function of excitation power under pulsed excitation at a sample temperature of 4.5~K.  The PL intensity was extracted from the area of Gaussian fits to the spectra. For the MoSe$_2$, trion and neutral exciton peaks were each fitted with a Gaussian, and the total PL intensity was calculated as the sum of the two areas. We clearly see that for low excitation power, the IEX PL dominates the spectrum, but the intralayer emission from the MoSe$_2$ layer increases more strongly with increasing excitation power, superseding the IEX emission. As the excitation power is increased further, the IEX PL shows clear signs of saturation, indicating that the increasing electron-hole pair density opens up the pathway for intralayer exciton recombination.
\subsection*{Supplementary note 6: correlation between IEX PL intensity and peak position}
\begin{figure}
\includegraphics*[width= 0.6 \linewidth]{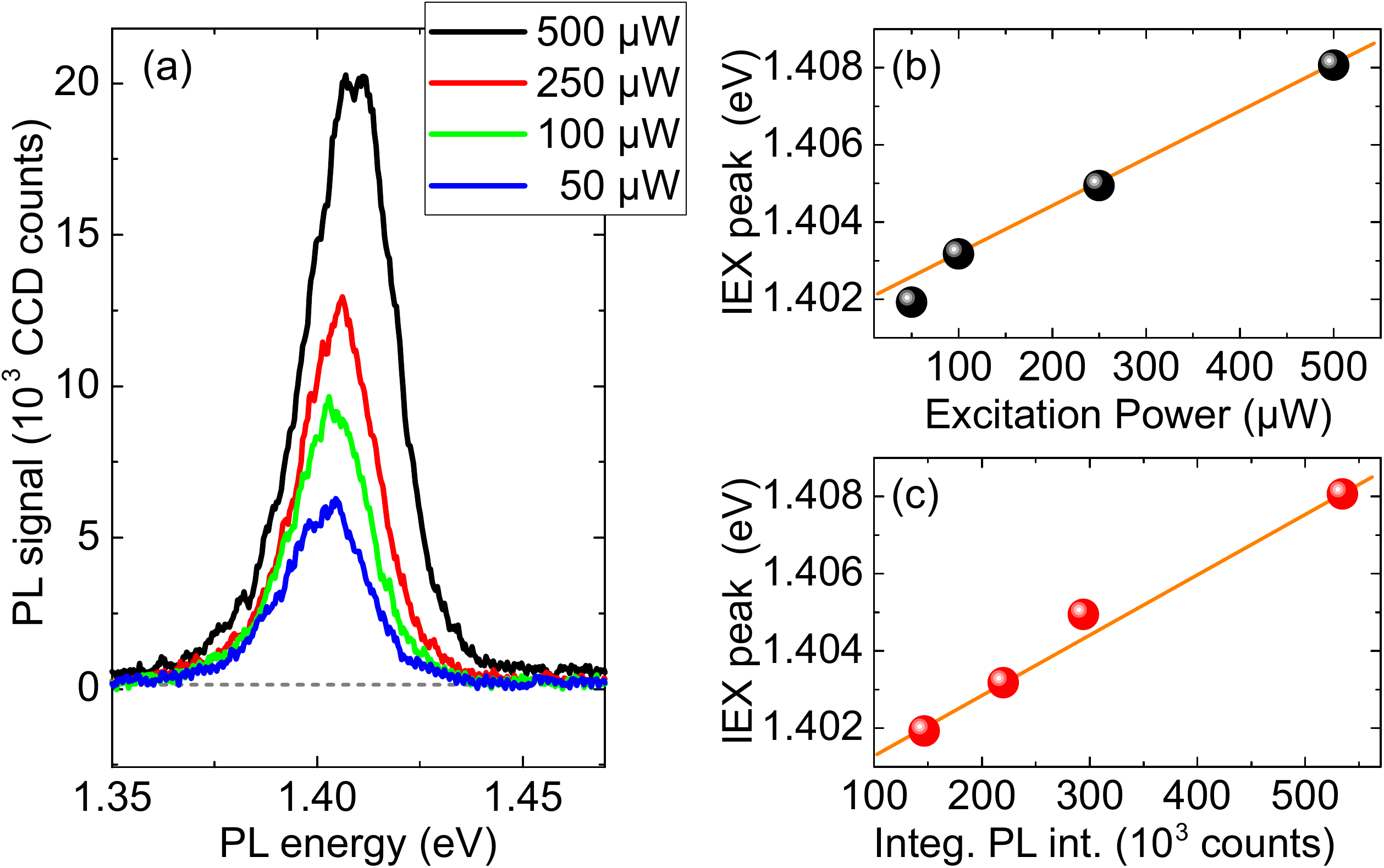}
\caption{\textbf{Dependence of IEX PL emission on excitation power and PL intensity} (a) PL traces extracted from TRPL measurements using 4 different excitation powers measured at 4.5~K. (b) IEX peak positions extracted from traces shown in (a) as a function of excitation power. (c) IEX peak positions extracted from traces shown in (a) as a function of integrated PL intensity. The orange lines in (b) and (c) serve as guide to the eye.}
\label{figSupp:Ex-Dens}
\end{figure}

In order to find the correlation between IEX PL peak position and exciton density, we analyzed a series of TRPL measurements, as outlined below.
Figure~\ref{figSupp:Ex-Dens}(a) shows a series of PL spectra extracted from TRPL measurements at 4.5~K using different excitation powers. The spectra were generated by averaging a 100~ps wide time window of time- and energy-resolved TRPL data, with the window starting at a time delay of 800~ps after arrival of the excitation laser pulse. We clearly observe a redshift of the PL peak position with decreasing PL intensity. For each spectrum, we determined the IEX PL peak position and the integrated PL intensity (area underneath the fit function) using a Gaussian. Plotting the IEX peak position as a function of excitation power (Fig.~\ref{figSupp:Ex-Dens}(b)) and integrated PL intensity (Fig.~\ref{figSupp:Ex-Dens}(c)) shows a near-linear increase of the IEX peak position as a function of either parameter, in stark contrast to the data shown in Fig.~3(b) of the main manuscript, which analyzes the \emph{time-integrated} PL data as a function of excitation power and clearly shows sublinear increase (note the logarithmic scale for the power axis in that figure), due to the reduction of the effective PL lifetime discussed in the main manuscript.
To generate the data shown in Fig.~5(f) of the main manuscript, we extracted time-averaged PL spectra at three different time delays for four different excitation powers, determined the  IEX PL peak position and the integrated PL intensity by fitting a Gaussian to the spectra, and normalized the intensities to the value obtained for the shortest time delay (800~ps) and the highest excitation power (500~$\mu$W).
\subsection*{Supplementary note 7: estimate of interlayer exciton density}
Here, we consider a mean-field approximation. As discussed by Laikhtman and Rapaport~\cite{Rapaport_PRB09}, this is valid in a regime where the interlayer exciton density $n_{IEX}$ is so small that the average exciton-exciton distance is substantially larger than the interlayer distance $d$: $n_{IEX}d^2\ll1$. Additionally, the exciton thermal wavelength $\lambda_{th}$ also needs to be smaller than the average exciton-exciton distance, $\frac{1}{\lambda_{th}^2}\gg n_{IEX}$. For a temperature of 4.5~K, we find $\lambda_{th}\approx10$~nm. Within this approximation, the blueshift $\Delta E_D$ induced by the dipolar exciton-exciton interaction is given by:
\begin{equation}\label{Shift}
\Delta E_D=\frac{4\pi n_{IEX} e^2 d}{\varepsilon_R \varepsilon_0}.
\end{equation}
Thus we find
\begin{equation}\label{niex}
n_{IEX}=\frac{\Delta E_D \varepsilon_R \varepsilon_0 }{4\pi  e^2 d}.
\end{equation}
Here, we utilize $\varepsilon_R$=4.5~\cite{Kumar20124627}, an interlayer distance $d=1$~nm, and a maximum energy shift $\Delta E_D=20$~meV, corresponding to the energy difference between the IEX peak positions under low-power cw excitation and high-power pulsed excitation. For the latter case, the energy value is extracted 800~ps after pulsed excitation to account for exciton diffusion into the potential minima, and therefore represents a conservative lower bound of the maximum blueshift. This yields a value of $n_{IEX}=4\cdot10^{10}$~cm$^{-2}$, which fulfills both of the limiting conditions mentioned above ($n_{IEX}d^2=4\cdot10^{-4}\ll1$, $\frac{1}{\lambda_{th}^2}=10^{12}$~cm$^{-2} \gg 4\cdot10^{10}$~cm$^{-2}$) for applying this simple \emph{Ansatz}.
This density is very low compared to the electron-hole pair density generated within the two constituent monolayers of the heterostructure under high-power pulsed excitation, which is on the order of $5\cdot10^{13}$~cm$^{-2}$, indicating a low efficiency of IEX formation under these excitation conditions. This low efficiency at high excitation powers is also reflected in the different power dependencies of intra- and interlayer exciton PL intensity, as shown in Fig.~\ref{figSupp:P-dep}.
\providecommand{\latin}[1]{#1}
\providecommand*\mcitethebibliography{\thebibliography}
\csname @ifundefined\endcsname{endmcitethebibliography}
  {\let\endmcitethebibliography\endthebibliography}{}

\end{document}